# Energy Levels of Odd-Odd nuclei Using Broken Pair Model

**I. M. Hamammu, S. Haq, J. M. Eldahomi**
*Physics Department, Faculty of Science, Garyounis University, Benghazi, Libya*

**ABASTRACT**

Simple methods in the frame work of the broken pair model have been worked out for odd-odd nuclei. The reliability of the model has been tested by reproducing the shell model results of limiting cases in which the broken pair model exactly coincides with the shell model. The model is then applied to calculate the energy levels of some nuclei in the Zirconium region. The model results compare reasonably well with the shell model as well as with the experimental data.

*Key Words : Shell model; Broken pair model; Approximation techniques.*

## INTRODUCTION

Although nuclear shell-model calculations have decisively benefited from the rapid increase in computing power, approximation techniques still play an important role in simplifying this issue (Elliot, 1990; Siiskonen, 2000; Papenbrock, 2004). Different approaches have been introduced to lower the dimensionality of the shell model space. The broken pair model (BPM) introduced earlier (Gambhir, 1969) is such an approximation. It has been successfully applied in the past to study even-even (Rimini, 1970; Gambhir, 1979, 1981) and even-odd nuclei (Hamammu, 1994, Haq, 1996). The aim of this work is to develop it for odd-odd nuclei. For this purpose we have derived the expressions for Hamiltonian matrix elements and E2 transition rates. The model is tested by comparing its results with those of the exact shell model (ESM) of Gloeckner (Gloeckner, 1975) where both models coincide. Using the same input data BPM has been applied to calculate the energy spectra of some nuclei in the Zr region. The model results were encouraging. The discrepancies wherever are owed to the crude assumption of the model. Better results are expected by enlarging the model space or extending the model by breaking more pairs.

**The Broken Pair Model:**

The corner stone in the BPM assumptions is the experimental fact that identical nucleons prefer to form pairs. The approximate BPM ground state for identical particle even nuclei is given as

$$|\phi_0\rangle = P^+ |0\rangle. \qquad (1)$$

where

$$a_p^+|0\rangle = \frac{1}{p}\left[\prod_a u_a^{j+\frac{1}{2}}\right]S_+^p|0\rangle. \qquad (2)$$

with p standing for the number of pairs and

$$S_+ = \sum_a \frac{\hat{a}}{2} Y_a A_{00}^+(aa). \qquad (3)$$

Being the pair distribution operator for angular momentum coupled pair operator $A_{00}^+(aa)$. Here $Y_a$ stands for the probability amplitude of pair distribution over the valence level a. The level a includes all the quantum numbers of single particle state except the projection m. $Y_a$ is calculated either by minimizing the Hamiltonian in the state (2) or by replacing it by

$$Y_a = \frac{u_a}{v_a} \quad \text{and} \quad u_a^2 + v_a^2 = 1. \qquad (4)$$

The parameters $u_a$ and $v_a$ are then determined by solving the BCS gap and number equations.

In our case the BPM basis states for odd nuclei in the first approximation are

$$|(r_p)\rangle = a_{r_p}^+ a_p^+|0\rangle. \qquad (5)$$

In the next higher approximations more $S_+$ operators are to be replaced by non pair creation operators $A_{JM}^+(ab)$ defined by

$$A_{JM}^+(ab) = \sum_{m_1,m_2}\begin{bmatrix} a & b & J \\ m_1 & m_2 & M \end{bmatrix}|am_1\rangle|bm_2\rangle. \qquad (6)$$

Here square bracket denotes the Clebsch Gordon coefficient (Brussard, 1977). This way by replacing all pair operators by non pair operators one gets the exact shell model space.

For a system of nonidentical nucleons having (2p+1) protons and (2n+1) neutrons, the first approximate broken pair state is written as angular momentum coupled product of the states (5)

$$|\Psi(r_p r_n JM)\rangle = \left[a_p^+ a_{r_p}^+ \otimes a_{r_n}^+ a_n^+\right]_M^J|0\rangle. \qquad (7)$$

**The Hamiltonian:**

The Hamiltonian of a system of identical nucleons (proton-proton ($H_{pp}$)/neutron-neutron $H_{nn}$) in second quantization is given by

$$H = \sum_{\alpha_1} \varepsilon^0_{a_1} a^+_{a_1} a_{a_1} + \frac{1}{4} \sum_{\alpha_1 \alpha_2 \alpha_3 \alpha_4} \langle \alpha_1 \alpha_2 | V | \alpha_4 \alpha_4 \rangle_A \, a^+_{a_1} a^+_{a_2} a_{\alpha_4} a_{\alpha_3}. \quad (8)$$

Where $\varepsilon^0_a$ being the single particle energy and the second term stands for antisymmetric normalized two body matrix elements. For neutron proton system, the interaction part $H_{np}$ is given by

$$H_{np} = \sum_{\alpha_p \alpha_p' \alpha_n \alpha_n'} \langle \alpha_p \alpha_n | V | \alpha'_p \alpha'_n \rangle \, a^+_{a_p} a^+_{a_n} a_{\alpha_n'} a_{\alpha_p'} \quad (9)$$

In BPM this Hamiltonian is recasted into a special form using the fermion anticommutaion rules The details are given in reference (Rimini, 1970; Gambhir, 1979, 1981 ) The reason for that is to express the matrix elements in terms of scalar products between BPM states for (2p+1) protons and (2n+1) neutrons. The method of evaluating matrix elements is also given in reference (Rimini, 1970; Gambhir, 1971,1981 ). The final form of the Hamiltonian is given by

$$H = H_o + H_{pp} + H_{nn} + H_{np}. \quad (10)$$

Here, $H_o$ contains no creation operators, $H_{pp}$ and $H_{nn}$ denote proton-proton and neutron-neutron parts of the Hamiltonian and $H_{np}$ contains two body part of the neutron-proton Hamiltonian. Their exact from is given in (Eldahomi, 1998).

This Hamiltonian when sandwiched between the states (7) gives the matrix elements of the Hamiltonian in terms of BPM overlaps. The $H_{np}$ matrix elements are given in the appendix, while the matrix elements of $H_o$, $H_{pp}$ and $H_{nn}$ are already given in (Gambhir, 1981).

## RESULT AND DISCUSSION

The model developed here has been applied to calculate the energy levels of nuclei in the Zr region. The model space was restricted to $2p_{1/2}$ and $1g_{9/2}$ on the proton side and $2d_{5/2}$ and $3s_{1/2}$ single particle states on the neutron side, assuming $^{88}$Sr as an inert core. The same configuration space had been used by Gloeckner (Gloeckner, 1975) to perform exact shell model calculations (ESM).

Firstly, the feasibility of the model has been tested by reproducing the shell model results of $^{90}$Y (Fig. 1), in which BPM and ESM configuration coincide exactly.

The model is then applied to study the energy spectra of other nuclei in the concerned region. The model results are shown (denoted by BPM) along with shell model results (whenever available) and experimental data (ENSDF) (denoted by EXPT) as well.

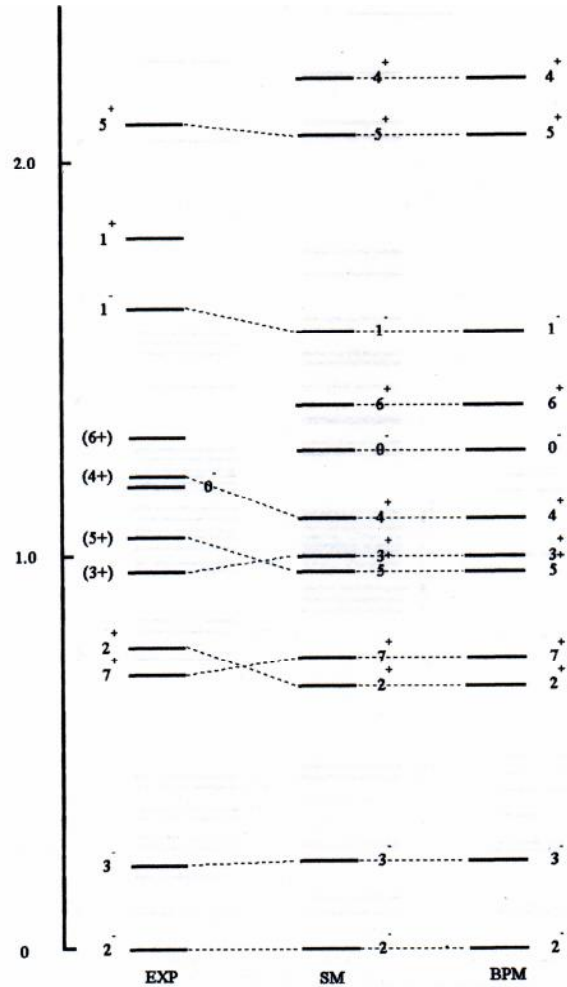

**Fig. 1. Experimental, shell model and broken pair model results of the energy levels for $^{90}$Y.**

**Nb Isotopes:**

Figs: 2 and 3 show the energy levels of $^{92}$Nb, $^{94}$Nb and $^{96}$Nb. The BPM results of $^{92}$Nb compares excellently well with ESM. For $^{96}$Nb, the BPM results compares very well with ESM as well as with EXPT. Although the results of $^{94}$Nb do not compare as expected, which may be owed to the crude assumptions of the model in which we take only the seniority $\nu=1$ state.

Energy Levels of Odd-Odd nuclei Using Broken Pair Model

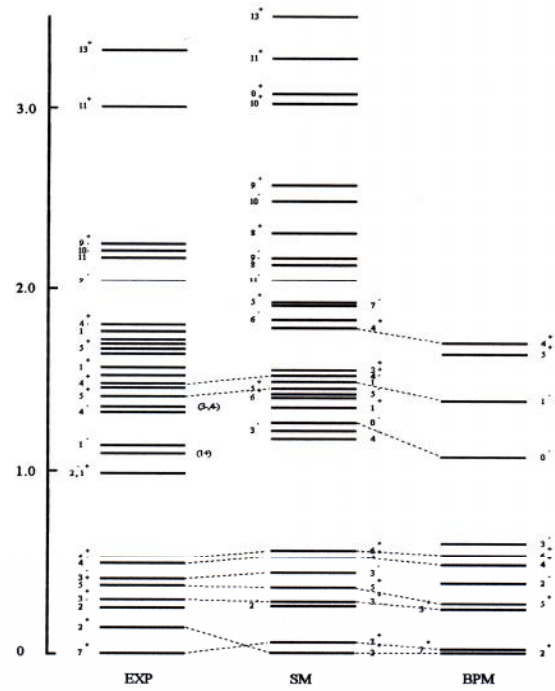

**Fig. 2. Experimental, shell model and broken pair model results of the energy levels for $^{92}$Nb.**

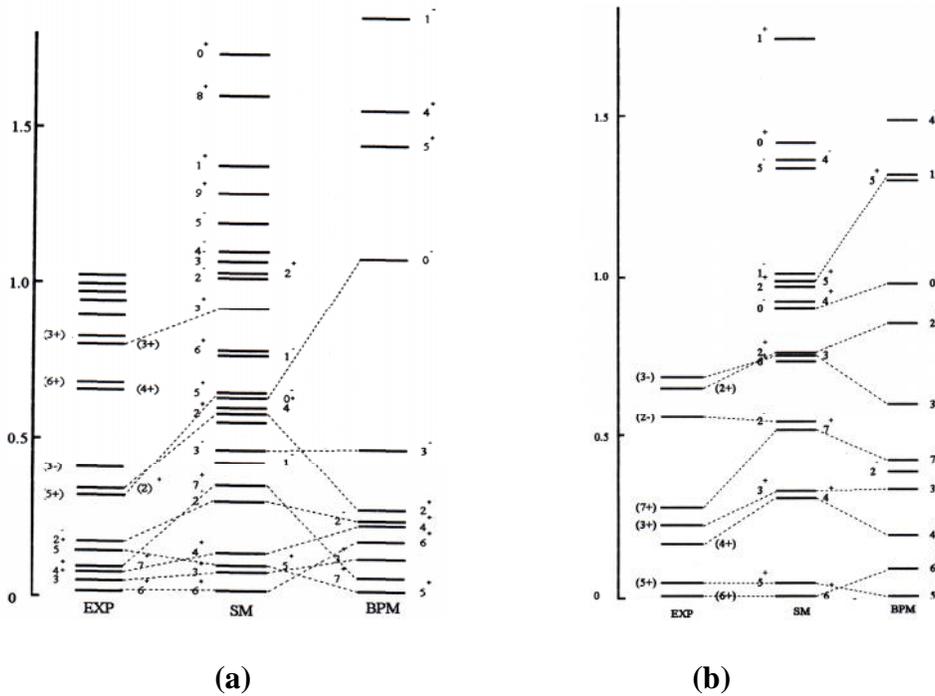

                (a)                                (b)

**Fig. 3. Experimental, shell model and broken pair model results of the energy levels for $^{94}$Nb (a) and $^{96}$Nb (b).**

**Y - Isotopes :**

The model results of $^{92}$Y and $^{94}$Y are shown in Figs(4a and 4b), which shows the quality of agreement between the BPM results and experimental data, the differences are owed to the fact that the choice of the two body matrix elements in which these nuclei were not included in fitting (Gloeckner, 1975). The experimental level $1^+$ in both cases either correspond to high seniority $\nu$ =3 neutron state, or due to the discarded $g_{7/2}$ level.

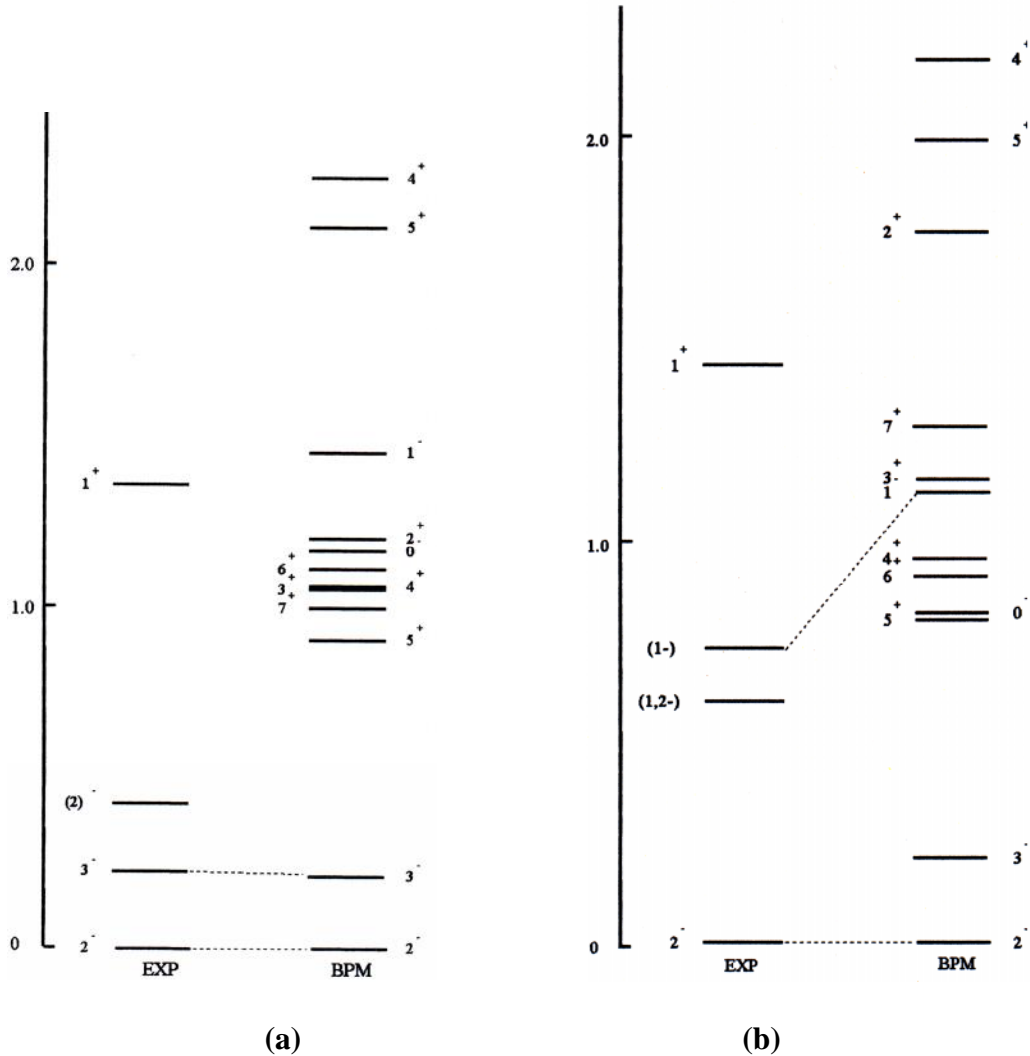

**Fig. 4. Experimental and broken pair model results of the energy levels for $^{92}$Y (a) and $^{94}$Y (b).**

**Tc - Isotopes :**

The calculated and experimental spectra of $^{94-98}$Tc are shown in Fig. 5. Theoretically the ground state spin is found to be $5^+$ for all Tc isotopes, while experimentally it is $7^+$ for $^{96}$Tc and $6^+$ for $^{98}$Tc. The other energy states compare reasonably well with the experimental data. A little discrepancy is found for negative parity states. The nonzero negative parity states are found by coupling $p_{1/2}$ proton level with $d_{5/2}$ neutron level. Hence if level $g_{7/2}$ is also included in the neutron space, the results of the negative parity states are expected to compare better with experimental data.

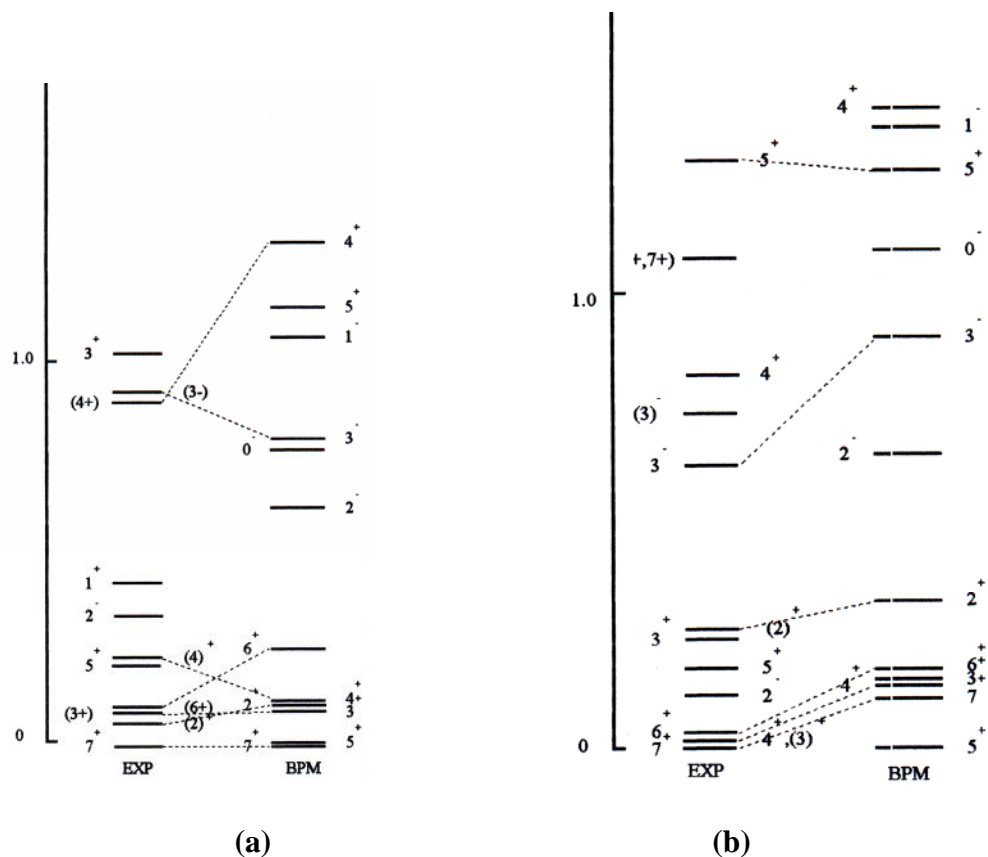

(a)                    (b)

**Fig. 5. Experimental and broken pair model results of the energy levels for (a) $^{94}$Tc and (b) $^{96}$Tc.**

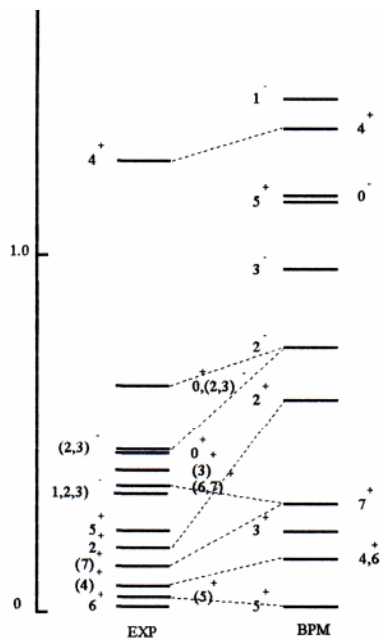

**Fig. 6. Experimental and broken pair model results of the energy levels for $^{98}$Tc.**

Energy Levels of Odd-Odd nuclei Using Broken Pair ModelRimini, A., & Weber, T. (1970). Number-conserving approximation to the shell model. *Physics Review*, C2 , 1573-1583

Siiskonen, T., & Lipas, P. O. (2000). Tests of shell-model truncation methods. *Physics Review* C61, 057302-1- 3
9